\documentclass[preprint,12pt]{elsarticle}
\usepackage{amssymb}
\usepackage{epsfig}
\usepackage{graphicx}
\usepackage{subfigure}
\usepackage{hyperref}
\usepackage{balance}
\biboptions{numbers,sort&compress}
\usepackage{epstopdf}

\begin{document}
\unitlength 1 cm
\newtheorem{thm}{Theorem}
\newcommand{\be}{\begin{equation}}
\newcommand{\ee}{\end{equation}}
\newcommand{\bearr}{\begin{eqnarray}}
\newcommand{\eearr}{\end{eqnarray}}
\newcommand{\nn}{\nonumber}
\newcommand{\vk}{\vec k}
\newcommand{\vp}{\vec p}
\newcommand{\vq}{\vec q}
\newcommand{\vkp}{\vec {k'}}
\newcommand{\vpp}{\vec {p'}}
\newcommand{\vqp}{\vec {q'}}
\newcommand{\bk}{{\bf k}}
\newcommand{\bp}{{\bf p}}
\newcommand{\bq}{{\bf q}}
\newcommand{\br}{{\bf r}}
\newcommand{\up}{\uparrow}
\newcommand{\down}{\downarrow}
\newcommand{\fns}{\footnotesize}
\newcommand{\ns}{\normalsize}
\newcommand{\cdag}{c^{\dagger}}

\title{ Quantum Phase transitions in the two dimensional ionic-Hubbard model}
\author{Ahmad Shahbazy}
\author{ Morad Ebrahimkhas}
\cortext[cor]{Corresponding author}
\ead{ebrahimkhas@iau-mahabad.ac.ir}
\address {Department of Physics, Mahabad Branch, Islamic Azad University,  Mahabad 59135, Iran}

\begin{abstract}
We employ the dynamical mean field approximation to study the effects of ionic potential ($\Delta$)
on the square lattice Hubbard model. 
At half-filling when the staggered potential ($\Delta$) dominates the on-site Hubbard interaction ($U$), the system is in the band insulator phase.
We find that competition between $U$ and $\Delta$ can suppress the gap to zero and  leading to an intermediate metallic region. 
At the large-$U$ limit, we identify a Mott insulator phase where the gap opens again and increases upon increasing the Hubbard interaction U.
For  $U\leq 9t$ and $\Delta\approx 0$ the phase of the system is metallic, but for  larger $U$ the system is in the Mott insulator phase.
\end{abstract}


\maketitle

\section{Introduction}
There are many quantum phase transitions which are driven by interactions.
The metal-insulator transition~\cite{ Georges-RMP, Jarrell} and semi metal-insulator transition~\cite{Ebrahimkhas1} serve as two examples. 
A famous model exhibiting competing interaction terms is 
the ionic Hubbard model (IHM), which includes a nearest-neighbour hopping, an on-site
repulsive interaction, and a staggered potential which separates neighbouring sites by
an energy shift~\cite{Lee}. When the staggered potential dominates, a band insulator results.
In a bipartite lattice,
the substrate can include a sub-lattice symmetry breaking (by ionic potential $\Delta$)
so the ground state of this system
will be a band insulator. The interaction can suppress the gap in the Band Insulators (BI) phase, by increasing repulsion interaction
the energy gap would be created again~\cite{Garg1}. On the strongly correlated limit,
when the Hubbard term dominates over the ionic term in IHM, i.e. $U\gg \Delta$,
the system goes to Mott insulator phase.
In this limit, charge fluctuation is frozen and no double occupancy is expected.
In higher dimensions, many theoretical and numerical analysis are published
which investigate  transition between Mott insulator and band insulator phases in the Ionic Hubbard Model (IHM). 
Many groups have applied  the dynamical mean field theory (DMFT)~\cite{Garg1, Japaridze},
detrimental quantum Monte carlo (DQMC)~\cite{Bouadim} and cluster dynamical mean field
theory (CDMFT)~\cite{Kancharla} on IHM to studying  the phase diagrams of mono layer
or bilayer structures~\cite{Ruger}.
We know that the real ground state of the 2D square lattice could be in anti-ferromagnetic (AF) phase~\cite{Byczuk}; These numerical results of the DMFT
equations for the IHM provides the existence of a critical interaction strength for the
transition from a correlated BI to a MI. They found that the ground state 
of the system is always insulating. Their results shown there
is a direct transition between the BI and the
AF-MI~\cite{Byczuk}. 
We present results on the two dimensional square lattice ionic Hubbard model obtained
by iterated perturbation theory (IPT) based DMFT~\cite{Garg1}. The DMFT  approach is one of the most
powerful methods to study strongly correlated systems. In this method an impurity problem should be solved: the interaction between a single site 
of a lattice hybridized to a bath; The bath of non-interacting electrons must be determined self-consistently.
 The DMFT is a powerful method~\cite{Georges-RMP} which
is capable of handling both the ionic and the Hubbard interaction term in IHM. This approximation becomes exact in limit of infinite coordination numbers.
For lower coordination number, the local self-energy (k-independent) becomes only an approximate description.
Therefore the most significant drawback of this approximation is expected to be underestimation of the spatial quantum fluctuations
~\cite{Ebrahimkhas1, Ebrahimkhas2, Jafari, Tran}, hence the values of the critical phase transition parameters maybe overestimated~\cite{Wu}.
But the overall picture emerging from a simple DMFT method is expected to hold even when more complicated methods are employed, such as
CDMFT \cite{Liebsch}. The DMFT were used in analysis the phase diagram of the Bethe lattice. 
The existence of metallic phase between the BI and MI
phase has been proved~\cite{Garg1}. The bilayer Hubbard model indicates a phase diagram similar to the IHM.  This is because a large inter-layer coupling results in a BI similar to the large staggered potential limit of the IHM~\cite{Okamoto}.
This problem could be generalized to the heterostructures such as 
$SrTiO3$ and $LaTiO3$, studies on these materials demonstrated a metallic phase appearing  at the interface between a MI and BI phase~\cite{Kancharla}.
In this manuscript, we first introduced the IHM model and DMFT 
technique with IPT impurity solver in section {\it II}, in Section {\it III} we present the results and discussions on the DMFT phase diagram in the $\Delta$-$U$ plane. We end this  manuscript by giving a conclusion in section {\it IV}.

\section{Model and Method}

The 2D ionic-Hubbard model on a bipartite (two sub-lattices A and B) square lattice
 is given by
\begin{eqnarray}
H=-t\sum_{i\epsilon A,j\epsilon B,\sigma}(c_{i\sigma}^{\dag}c_{j\sigma}+h.c.)
-\Delta \sum_{i\epsilon A}{n_i}+\Delta \sum_{i\epsilon B}{n_i}\nn\\ 
+U\sum_{i}n_{i\uparrow}n_{i\downarrow}
-\mu \sum_{i\sigma}n_{i}
\label{IHH}
\end{eqnarray}
where $t$ denotes the nearest neighbour hopping and  $c_{i\sigma}^{\dag}(c_{i\sigma})$
are the creation (annihilation) operator of electrons at site $i$ with spin $\sigma$.
$\Delta$ is the ionic staggered potential which  alternates sign between sites in sub-lattice
$A$ or $B$.
 The number operator  is $n_{i,\sigma}$ which determines the number of electrons at site $i$
  with projection of spin  $\sigma$. The chemical potential is $\mu=U/2$ at half filling
  and $U$ designates the on-site electron-electron repulsion
Hubbard potential. We take $t$ as the energy unit through out this paper,
  we will consider the average filling factor $\frac{\langle n_A\rangle +\langle n_B \rangle}{2}=1$.
  
  In this model for $t>0$ and non-interacting limit $U=0$, a band insulator phase represents
  with energy gap, $E_{gap}=2\Delta$. The sites $A(B)$ have potential $-\Delta(\Delta)$
  , respectively. The electrons prefer to doubly occupy lower bands Fig.{\ref{BI.fig}.
  The average filling factors at half-filling is $\langle n_A\rangle=2$, $\langle n_B\rangle=0$ 
  at $U=0$ and $t=0$.

 \begin{figure}[tb]
  \begin{center}
    \includegraphics[width=5cm,height=4cm,angle=0]{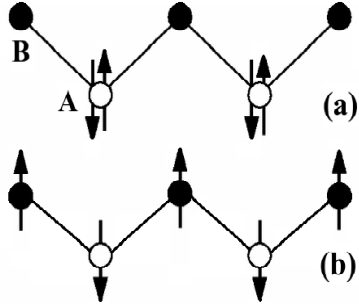}
    \caption{(Color online) The comparison between electrons occupancy on sites $A$ and $B$ 
    in BI and MI phase. The $A$ sites have lower potential than $B$ sites.
    $(a)$ In band insulator the  electrons choose doubly 
    occupy lower bands. $(b)$ In Mott insulator phase the electrons prefer to occupy 
    only one sites, $A$ and $B$.}
     
    \label{BI.fig}
  \end{center}
\end{figure}

In other limit, for $U\gg\Delta$, the system goes to MI phase with
 $\langle n_A\rangle=\langle n_B \rangle=1$~\cite{Hafez}. The flow equation methods predicted in the intermediate region
 at some energy scales the ionic and Hubbard potentials can cancel each other's effect ~\cite{Hafez}.
 From this point of view the tight-binding term dominates the nature of the ground state. Therefore we expect the intermediate
 phase to be a metal~\cite{Garg1,Hafez}. Here we study the IHM using the DMFT approach~\cite{Garg1}.
 To formulate the DMFT machinery, first we consider IH Hamiltonian on the square lattice in
 $\it{paramagnetic~phase}$.
 The interaction Green's function in the bipartite lattice are,

\begin{equation}
G^{-1}(\vk, \omega^+) =
\left(
\begin{array}{ll}
\zeta_{A}(\vk, \omega^+) & -\epsilon(\vk)\\
 -\epsilon(\vk)           &\zeta_{B}(\vk, \omega^+) \\
   \end{array}
   \right)
\label{IGF}
\end{equation}
where $\vk$ is the momentum vector in first Brillouin zone (FBZ), 
$\epsilon(\vk)=\epsilon(k_x, k_y)=-2t(cos(k_x)+cos(k_y))$ is the energy dispersion
for square lattice, and $\zeta_{A(B)}=\omega^+ \mp \Delta + \mu +\sum_{A(B)}(\omega^+)$
with $\omega^+=\omega +i0^+$. The self-energy, $\sum_{\alpha}(\omega^+)$ in DMFT is local and 
independent of $\vk$. The matrix elements of the self energy are diagonal and the
off-diagonal elements are zero.  The local Green's function of two sub-lattices 
could be written as, $G_{\alpha}(\omega^+)=\sum_{\vk}G_{\alpha \alpha}(\vk, \omega^+)$,

\begin{equation}
G_{\alpha}(\omega^+)=\zeta_{\bar{\alpha}}(\omega^+)\int_{-\infty}^{\infty}d\epsilon
\frac{\rho_0(\epsilon)}{\zeta_A(i\omega_n)\zeta_B(i\omega_n)-\epsilon^2}
\label{LGF}
\end{equation}

where for $\alpha=A(B)$, $\bar{\alpha}=B(A)$. The bare density of state obtained for
the square lattice,
\begin{eqnarray}
\rho_0(\epsilon)&=&\frac{1}{2t\pi^2}\int_{0}^{\pi}dk_{x}\int_{0}^{\pi}dk_y
\delta(\tilde{\epsilon} +cos(k_x)+cos(k_y))\nn\\
&=&\frac{1}{2t\pi^2 }\int_{-1}^{1}d{x}\int_{-1}^{1}dy\frac{\delta(\tilde{\epsilon}+x+y)}
{\sqrt{(1-x^2)(1-y^2)}}\nn\\
&=&\frac{1}{2t\pi^2 }\int d{x}\frac{1}{\sqrt{(1-x^2)(1-(\tilde{\epsilon}+x)^2)}}
 \label{DOS.EQ}
\end{eqnarray}

where $\tilde{\epsilon}=\epsilon/2t$. The Eq.~\ref{DOS.EQ} can be evaluated numerically
for these two intervals, $-1\leq x\leq 1$ and
$-1\leq (\tilde{\epsilon}+x)\leq 1$. We first guess the initial filling factor and
 self-energy~\cite{Garg1}. Then we determine the host Green's function from the Dyson's equation
$\mathcal{G}_{0\alpha}^{-1}(\omega^+)=G_{\alpha}{-1}(\omega^+)+\sum_{\alpha}(\omega^+)$. 
Afterwards, we solve impurity problem and obtain
 $\sum_{\alpha}(\omega^+)=\sum_{\alpha}(\mathcal{G}_{0\alpha}(\omega^+))$. We use IPT method as impurity
 solver~\cite{Garg1}. The iteration in these steps continues until convergence is reached.
 We  calculate the density of state by
  $\rho_\alpha(\omega)=-\sum_{\vk}\Im [G_{\alpha}(\vk,\omega^+)]/\pi$. From the particle-hole symmetry
  at half-filling we obtain $\rho_A(\omega)=\rho_B(-\omega)$ for the $DOS$ of two sub-lattices.
  The total $DOS$ for square lattice obtains via $\rho(\omega)=\rho_A(\omega)+\rho_B(\omega)$.

\section{Results and Discussions}
\subsection{DOS}

In Fig.~\ref{HBand.fig} has been showed the DOS for some values of $U$ at a constant $\Delta=0.3t$. 
This figure can covers the whole range of energies. For small values of $U$ ($U \le U_{c1}$) 
the spectrum has simple gap. By increasing $U$ ($U_{c1}\le U\le U_{c2}$), the overall 
feature of the low energy spectrum  changes and
the DOS around $\omega=0$ has singularity and the gap was closed.
 When  $U$ increases ($U\ge U_{c2}$) we see the upper and lower Hubbard band appear, symmetrically.
The formation of the Hubbard bands are shown in Fig.~\ref{HBand.fig}.

\begin{figure}[htbp]
  \begin{center}
    \includegraphics[width=7cm,height=6cm,angle=0]{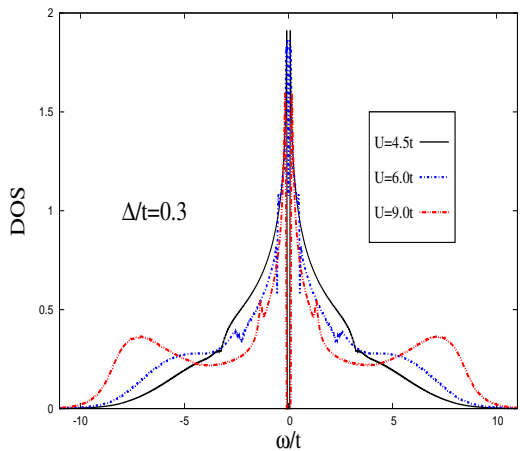}
    \caption{(Color online) The density of states for square lattice  in IHM
    are plotted for $U=4.5t,~6.0t,~9.0t$, $ \Delta=0.3t$. In this figure, the Hubbard bands
    appeared symmetrically in both sides of $\omega=0$.
        } 
    \label{HBand.fig}
  \end{center}
\end{figure}

The DOS of square lattice in Hubbard model ($\Delta=0$) has Van-Hov singularity around the Fermi
level. When the ionic potential ($\Delta$) was added, at $U=0$ the energy gap is
 $E_{gap}\simeq2\Delta$, and two singularities was appear in both sides of the gap.
 By increasing $U$ in $U\le U_{c1}$ interval the Van-Hov singularities moves towards lower energies.
 In $U\ge U_{c2}$ interval singularities move towards higher energies. 
 
 \subsection{$E_{gap}$}
 The energy gap could be calculated in two methods, first through measurement the gap of DOS, second
 from the formula of $E_{gap}$,
 
 \begin{equation}
   E_{gap}=Z|\Delta -U\delta n/2+S|
   \label{Egap.eqn}
\end{equation}

Where $Z$ is independent of $\alpha$, $\delta n=\frac{\langle n_A\rangle -\langle n_B \rangle}{2}$
and $S=P\int_{-\infty}^{\infty} d\omega \sum''_A(\omega)/\pi \omega$. We could find the gap strongly 
depends on repulsion  interaction~\ref{Egap.fig}.  
In Fig.~\ref{Egap.fig}, the dots diagram obtained from measurement the energy gap from
the DOS and the open square with error
bars obtained from Eq.~\ref{Egap.eqn} for $\Delta=0.5t$. The inner figure is plotted for $\Delta\simeq 0$. We found that at $\Delta = 0$ the gap is opened in $U>U^{0}_{c}$ interval ($U^{0}_{c}\geq 9t$).
This is not in accordance with the results of Ref.~\cite{Bouadim} in $\Delta\simeq 0$ which used
determinant quantum Monte Carlo (DQMC)  method. 
As can be seen the actual (interacting) gap strongly depends on the interaction parameter
U in BI and MI phases.
The DOS and error bars diagrams have excellent agreement with each other in both figures.

\begin{figure}[htbp]
  \begin{center}
    \includegraphics[width=7cm,height=5cm,angle=0]{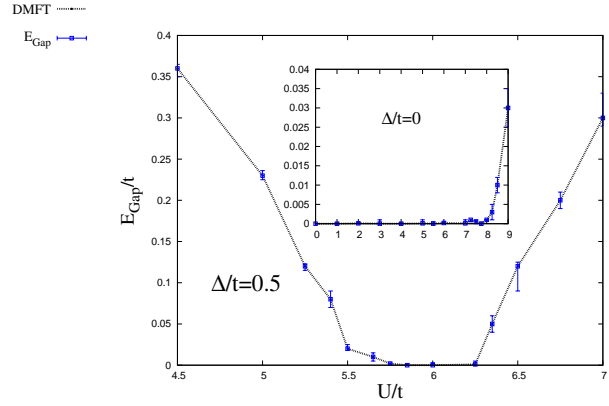}
    \caption{(Color online) The energy gap for $\Delta/t=0.5$ with two methods plotted
    $vs$ $U$  in unit of $t$. The gap (dots) is plotted using the DOS. The results with error bars
    (open square) obtained from Eq.~\ref{Egap.eqn}. The error bars related to numerical errors
     in calculation of $\sum''(\omega)$ at small $\omega$~\cite{Garg1}. The inner figure shows
     $E_{gap}$ in $\Delta/t=0$. 
        } 
    \label{Egap.fig}
  \end{center}
\end{figure}

The  error bars  estimated from calculation of 
$\sum ''(\omega \rightarrow 0 )$. We increased the accuracy of the calculations and we
used accurate methods in numerical integration. These error bars  obtained  by considering
these two results from two integration methods.

\subsection{Self-Energy}

For more Discussions in the scale of energy gap we consider the self energy of this system, 
 $\sum_{\alpha}(\omega^+)=\sum'_{\alpha}(\omega)+i\sum''_{\alpha}(\omega)$, where shape of imaginary 
 part of self energy can show insulating or metallic phase of systems~\cite{Garg1,Ebrahimkhas2}.
 The $\sum''_{A}(\omega)$ vanishes around $\omega=0$ in BI and MI phases, according to the
 Fermi liquid theory in metallic phase $\sum''_{A}(\omega)=\omega^2$~\cite{Garg1}.
 We showed these results in Fig.~\ref{Self.fig}.

\begin{figure}[htbp]
  \begin{center}
    \includegraphics[width=8cm,height=7cm,angle=0]{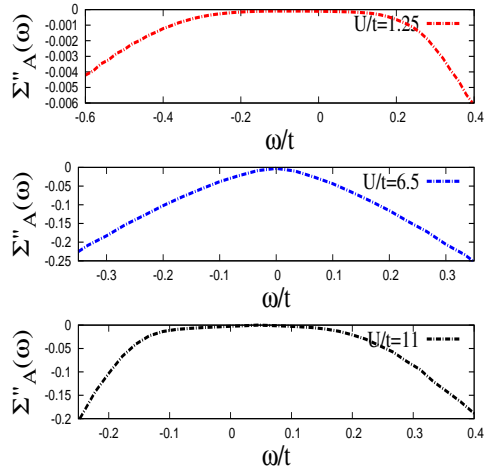}
    \caption{(Color online) The Imaginary part of self energy for $\Delta/t=0.1$ plotted
    $vs$ $\omega$. The top and low panel are related to Band insulator and Mott insulator.
    The middle panel is related to Metallic phase. 
        } 
    \label{Self.fig}
  \end{center}
\end{figure} 

\subsection{Phase~diagram}

By repeating the calculation for wide range of Hubbard potential and ionic potential,
we can map out the phase diagram of the square lattice for ionic-Hubbard model.
When $\Delta=0$, IHM is reduced to the Hubbard model. As it is known~\cite{schafer}
for the Hubbard model on the square lattice for finite values of $U$ according to the DMFT, phase 
transition  takes place from metallic to MI phase.
We could find interesting results different from phase diagram of 
honeycomb lattice. The phase of the honeycomb lattice in $\Delta=0$ and finite $U$,
is semi-metal~\cite{Ebrahimkhas2}. In $\Delta>>0$ region 
the metallic region shrinks to a single metallic point for honeycomb lattice.
But in the 2D square lattice for $\Delta\approx0$, the only critical repulsion potential
is $U_{c}\approx 9.1t$. In the range of $U\leq 9.1t$, the metallic phase is dominate. 
By increasing the amount of $U$ in $\Delta\approx0$ the MI phase
is overcome~\ref{DOS.fig}.

\begin{figure}[htbp]
  \begin{center}
    \includegraphics[width=8cm,height=7cm,angle=0]{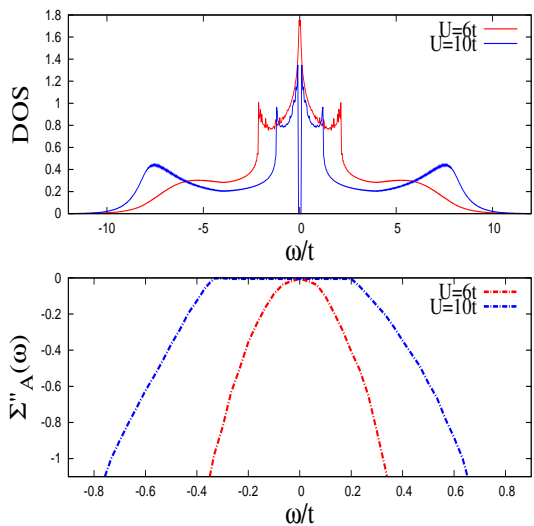}
    \caption{(Color online) The DOS and imaginary part of self energy
    of square lattice for $\Delta/t=0$ and $U/t=6~,10$ are plotted
    $vs$ $\omega$. The top and low panel are related to DOS and $\sum''_{A}$ respectively.
        } 
    \label{DOS.fig}
  \end{center}
\end{figure} 

It is well known that, phase transition from metallic to insulating phase for
Hubbard model ($\Delta=0$) is the first order. We could find coexistence region
of metallic and insulating phase at finite $\Delta$, last Ref. in ~\cite{schafer}.
In the square lattice a little increasing in $U\simeq9.1t$ for $\Delta \approx 0$ can open the gap,
this could be explained as the easiness  the phase transition of Schrödinger electrons
compared to Dirac electrons~\cite{Ebrahimkhas1}. 
If we employ the DMFT to the IHM at
zero temperature allowing for spontaneous AF long-range
order, We find that the ground state is always insulating phase
with a gap in DOS. We could find a direct transition between the BI and the
AF-MI~\cite{Byczuk}.
In Ref.~\cite{Bouadim} at $\Delta=0$ we could
not see the phase transition and the system is in Mott insulator phase.
But DMFT gives metallic phase for finite $U$ which is the mean-field artefact.
The inclusion of inter-site magnetic fluctuations  could be yield an anti-ferromagnetic 
insulating phase for all U (in DQMC method) on the two dimensional systems.

But our results are obtained in the paramagnetic phase as ground state. One can
expect that the metallic phase will survive if AF is suppressed due to frustration, in
the same way as in the DMFT treatment in the $\Delta = 0$
limit~\cite{Georges-RMP}. In future work we could consider a model in
which such frustration is explicitly included.

\begin{figure}[htbp]
  \begin{center}
 \includegraphics[width=7cm,height=5cm,angle=0]{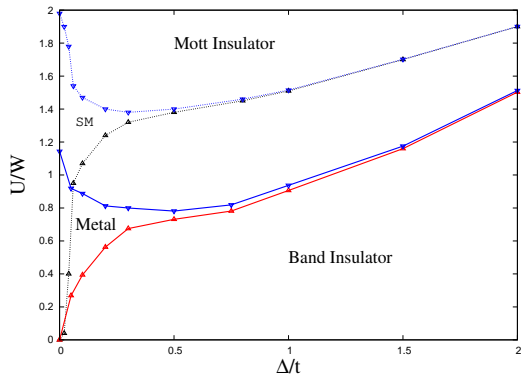} 
    \caption{(Color online) The phase diagram of IHM on the 
 square lattice by DMFT(IPT) method. The metallic phase is between two insulator phases.
 We plotted the phase diagram of honeycomb lattice in competition with square lattice. 
        } 
    \label{PD.fig}
  \end{center}
\end{figure} 

Our results in prediction of phase transition in $\Delta\approx0$ have nearly good agreement with results 
obtained from other DMFT methods on Hubbard model ($\Delta=0$)~\cite{schafer}.
For IHM on the square lattice at finite temperature~\cite{Martinie} and on Bethe lattice at $T=0$
~\cite{Garg1} show that for $\Delta\approx0$ and $\Delta\ll1$ phase transition from metallic to MI
phases take place.
 In this paper, three phases  BI, MI and Metal are predicted at zero temperature $T=0$. The w is bandwidth.
But if the temperature increased $e.g.$ $ k_{B}T=0.01$, the phase diagram will be changed~\cite{Martinie}.
In Ref.~\cite{Martinie} for $\Delta \approx 0$ and $0<U<U_{c1}\simeq2$ the 2D square lattice is in
Metallic phase. Similar studies have been done on bilayer square lattice IHM with interesting 
results~\cite{Jiang}. 
For large $\Delta$ we can consider that the metallic region goes to the narrow metallic region, which
has good agreement with~\cite{Martinie}.

Our results in small $\Delta$ have good agreement with Ref.~\cite{schafer} which 
used Hubbard model. The changes of regions in phase diagrams of the systems in
$\Delta\geq1$ could be compared with~\cite{Martinie} that the metallic region becomes narrow.
We plotted the phase diagram of honeycomb lattice in Fig.~\ref{PD.fig}. The two systems have
same treatment in quantum phase transition, but the middle phase is semi-metal(SM) in the honeycomb lattice.

\section{Conclusion}
We studied the ionic-Hubbard model on 2D square lattice
by IPT impurity solver. We calculated density of state, energy gap and self-energy at $T=0$ conditions.
Our results show a metallic phase is between to insulator phases. For $\Delta>0$
the system is in band insulator ($0<U<U_{c1}(\Delta)$). By increasing $U$  in
 $U_{c1}(\Delta)<U<U_{c2}(\Delta)$ region, the system shows metallic character. Finally, for 
 $U>U_{c2}(\Delta)$ the correlations transform the metallic phase to Mott Insulator.
 The calculations showed by increasing $\Delta$ the metallic region becomes narrow.

\section{Acknowledgements} 
We would like to thank M. Hafez Torbati for many useful discussions.

\end{document}